\documentclass[aps,prl,twocolumn,superscriptaddress,showpacs]{revtex4-1}
\usepackage{amsmath,amssymb}
\usepackage{color,framed}
\usepackage[pdftex]{hyperref,graphicx}
\usepackage{xspace}
\usepackage{hypcap}
\usepackage{dsfont}
\hypersetup{colorlinks = true, urlcolor = blue, linkcolor = blue, citecolor = blue}

\begin{document}
\title{Conductance spectroscopy of nontopological-topological superconductor junctions}

\author{F.~Setiawan}\email{setiawan@umd.edu}
\author{William~S.~Cole}
\author{Jay~D.~Sau}
\author{S.~Das~Sarma}
\affiliation{Condensed Matter Theory Center, Station Q Maryland, and Joint Quantum Institute,
Department of Physics, University of Maryland, College Park, Maryland 20742, USA}
\date{\today}

\begin{abstract}

We calculate the zero-temperature differential conductance $dI/dV$ of a voltage-biased one-dimensional junction between a nontopological and a topological superconductor for arbitrary junction transparency using the scattering matrix formalism. We consider two representative models for the topological superconductors: (i) spinful $p$-wave and (ii) $s$-wave with spin-orbit coupling and spin splitting. We verify that in the tunneling limit (small junction transparencies) where only single Andreev reflections contribute to the current, the conductance for voltages below the nontopological superconductor gap $\Delta_s$ is zero and there are two symmetric conductance peaks appearing at $eV = \pm \Delta_s$ with the quantized value $(4-\pi)2e^2/h$ due to resonant Andreev reflection from the Majorana zero mode. However, when the junction transparency is not small, there is a finite conductance for $e|V| < \Delta_s$ arising from multiple Andreev reflections. The conductance at $eV = \pm \Delta_s$ in this case is no longer quantized. In general, the conductance is particle-hole asymmetric except for sufficiently small transparencies. We further show that, for certain values of parameters, the tunneling conductance from a zero-energy conventional Andreev bound state can be made to mimic the conductance from a true Majorana mode.

\end{abstract}

\maketitle

\newcommand{\beq}{\begin{equation}}
\newcommand{\eeq}{\end{equation}}
\newcommand{\ba}{\begin{align}}
\newcommand{\ea}{\end{align}}
\newcommand{\nl}{\nonumber\\}
\newcommand{\ssc}{$s$SC\xspace}
\newcommand{\psc}{$p$SC\xspace}
\newcommand{\bmat}{\begin{matrix}}
\newcommand{\emat}{\end{matrix}}
\newcommand{\sgn}{\operatorname{sgn}}
\newcommand{\jtle}{\widetilde{j}_{NL,\nu}^e}
\newcommand{\jtlh}{\widetilde{j}_{NL,\nu}^h}
\newcommand{\jtre}{\widetilde{j}_{NR,\nu}^e}
\newcommand{\jtrh}{\widetilde{j}_{NR,\nu}^h}
\newcommand{\iinpl}{\widetilde{j}^{\mathrm{tot}}_{NL,\nu}}
\newcommand{\iinpr}{\widetilde{j}^{\mathrm{tot}}_{NR,\nu}}

Topological superconductors (TSs), which host localized Majorana zero modes (MZMs) at their boundaries or in defects, have drawn a considerable amount of interest as the most promising platform for topological quantum computation~\cite{Alicea12,Beenakker,Leijnse,Tudor,Franz,Sarma,beenakker16}. One straightforward experimental signature of the MZM is the zero-bias conductance peak~\cite{Tanaka95,Tanaka00,Sengupta01,Law09,Flensberg10,Wimmer11,jay10,Setiawan15} (robustly quantized at a value $2e^2/h$) of a normal metal-superconductor (NS) junction. This quantized conductance arises from perfect Andreev reflection facilitated by the MZM at the edge of the TS \cite{Sengupta01,Law09,Flensberg10,Wimmer11,Setiawan15}. A recent group of proposals for realizing TS in realistic solid state systems \cite{Fu08,roman,oreg,jay10,alicea} led quickly to several suggestive experimental observations of zero-bias conductance peaks \cite{Mourik,deng12,Das,Churchill,Finck,Perg,Jia16,Zhang16}. While such features have been carefully shown to correspond to the topological regime, the observed zero-bias conductance is still far below the expected quantized value. This deviation can be attributed at least in part to thermal broadening in the normal metal lead which in turn broadens the zero-bias peak and reduces its maximum conductance value. 

Since the effect of thermal broadening in a superconductor is exponentially suppressed by the superconducting gap $\Delta_s$, Peng \textit{et al}.~\cite{Peng15} have proposed to use a conventional superconducting lead as the ``probe" in an MZM tunneling experiment. They found that, in the tunneling limit (or small junction transparency), the MZM manifests as two symmetric conductance peaks quantized at $G_M = (4-\pi)2e^2/h$ appearing at the threshold voltages $eV = \pm \Delta_s$, i.e., when the BCS singularity of the probe lead aligns with the MZM. The result was derived using the nonequilibrium Green's function approach in the perturbative limit where contributions to the current come only from direct tunneling and single Andreev reflections.

In this Rapid Communication, we address the issue of whether this quantization is robust beyond the perturbative limit. To this end, we use a scattering matrix approach which easily incorporates multiple Andreev reflection (MAR) processes~\cite{KBT,Averin,Bagwell,suppl}. In the limit of small transparencies, where the dominant contribution to the current comes only from single Andreev reflections, our results agree with those of Refs.~\cite{Peng15,Yeyati}. However, when MAR gives a finite contribution to the current, we find that the MZM conductance quantization breaks down. Moreover, we also show that the conductance peak of a zero-energy Andreev bound state (ABS) may look similar to that of an MZM, such that (within realistic experimental resolution) the MZM and zero-energy ABS conductances may not be distinguishable from each other.

\begin{figure}[h!]
\capstart
\begin{center}
\includegraphics[width=\linewidth]{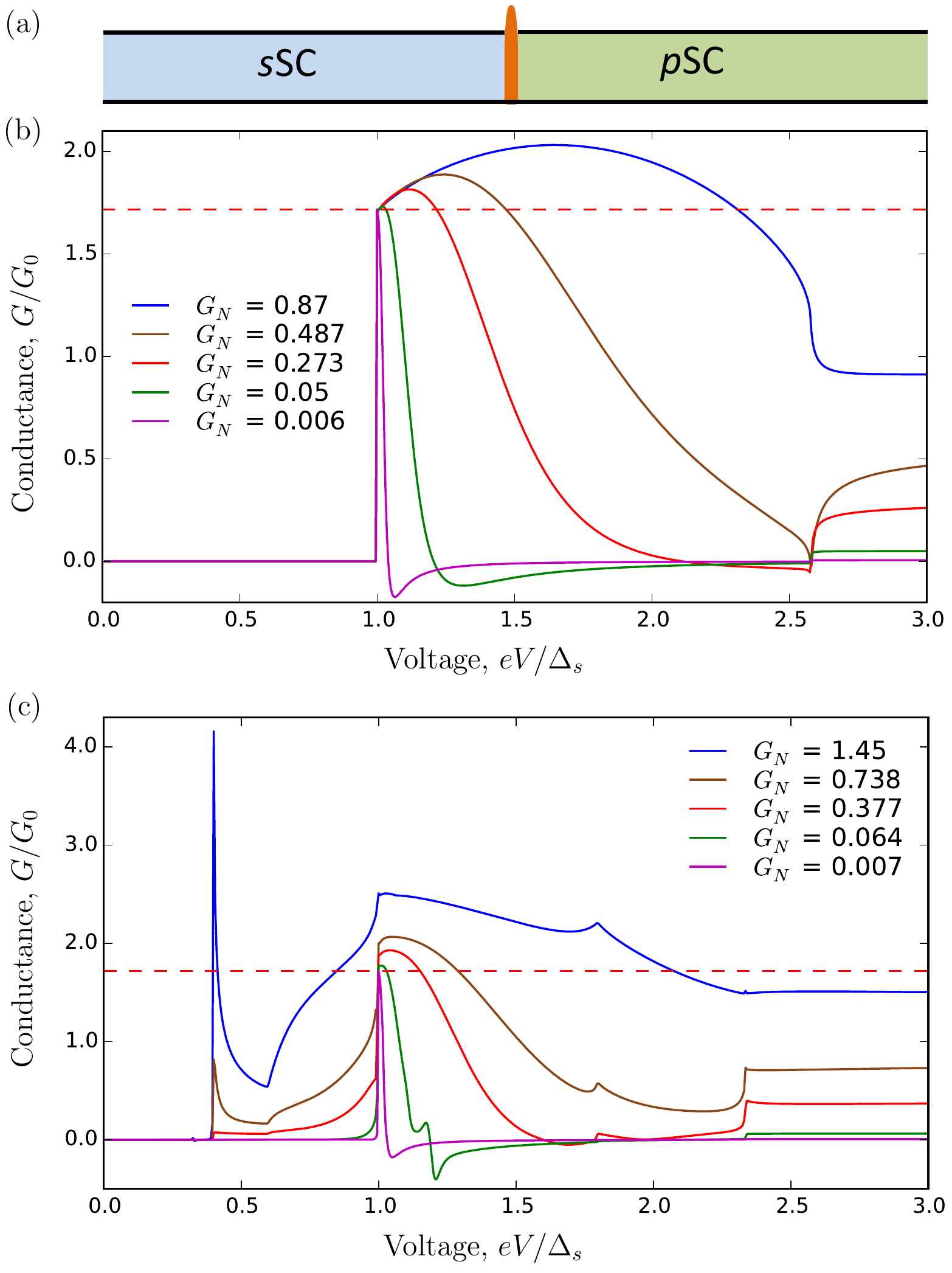}
\end{center}
\caption{(color online) (a) Schematic diagram for an \ssc-\psc junction with a delta barrier in the middle. (b),(c) Normalized differential conductance $G/G_0$ vs bias voltage $V$ for an \ssc-\psc junction with different Zeeman fields: (b) $V_Z =$ 40 K ($V_Z = 2\mu_p$) and (c) $V_Z =$ 22 K ($V_Z = 1.1\mu_p$). We present the results for several junction transparencies $G_N$ where $G_N$ is the differential conductance at high voltages which is the conductance of the corresponding NN junction. All conductances are expressed in units of $G_0 = e^2/h$. The red dashed line [$G_M= (4-\pi)2e^2/h$] is the conductance value of tunneling into the MZM. In the tunneling limit, the conductance displays a step jump from $G= 0$ to $G = G_M$ at the threshold voltage $eV = \Delta_s$. In all plots, we used the following parameters: $m =$ 0.015$m_e$, $\mu_s =$ 200 K, $\mu_p =$ 20 K, $\Delta_s$ = 2.5 K, and $\Delta_p$ = 0.039 eV$\cdot$\AA. The topological gaps of the \psc are (b) $\Delta_{\mathrm{topo}} = 4$ K and (c) $\Delta_{\mathrm{topo}} = 2$ K, $3.4$ K.} \label{fig1}
\end{figure}

We begin by modeling a one-dimensional superconductor-normal metal-superconductor (SNS) junction where one of the superconductors is nontopological (i.e., conventional $s$-wave) and the other is topological, with a delta-function barrier of strength $Z$ separating them [Fig.~\ref{fig1}(a)]. We calculate the conductance ($G = dI/dV$) for the SNS junction using the scattering matrix formalism as detailed in Ref.~\cite{suppl}, in complement to the Green's function approach commonly employed in the literature to study the TS junctions~\cite{Peng15,Yeyati,aguado,Denis}. In this formalism, the scattering processes  are partitioned into scattering processes at the left NS interface, tunnel barrier and right NS interface. We have generalized the formalism to a general superconductor-superconductor junction where one needs to calculate only the scattering matrices at the left and right NS interfaces~\cite{BTK,Setiawan15}. These scattering matrices can be computed easily using the numerical transport package Kwant~\cite{kwant,suppl}. The junction transparency in this model is modified by tuning the delta function barrier strength $Z$. Since there are also effective barriers arising from the sharp variation of model parameters across the junction, e.g., Fermi level mismatch, spin-orbit coupling, etc., we use a parameter-independent measure, $G_N$, to characterize the junction transparency, where $G_N$ is the conductance of the SNS junction at high voltages which is the conductance of the corresponding NN junction. We note that since the power dissipated  $IV$ by an SNS junction is always non-negative, it follows that the current for an SNS junction is always non-negative for positive bias voltage~\cite{suppl}.

In this Rapid Communication, we examine two models for the topological superconductor: (i) a spinful $p$-wave superconductor (\psc), and (ii) a spin-orbit-coupled superconducting wire (SOCSW). First, let us consider the junction between an $s$-wave superconductor (\ssc) and a spinful \psc, shown schematically in Fig.~\ref{fig1}(a).
The Hamiltonian of the system can be written in the Bogoliubov-de Gennes (BdG) form
\begin{equation}
H_j(x) = \frac{1}{2} \int dx \Psi_j^\dagger(x) \mathcal{H}_j \Psi_j(x),
\end{equation}
where $\Psi_j(x) = \left( \psi_{j\uparrow}(x), \psi_{j\downarrow}(x), \psi_{j\downarrow}^\dagger(x),-\psi_{j\downarrow}^\dagger(x) \right)^\mathrm{T}$ are Nambu spinors with $\psi^\dagger_{j\sigma}(x)$ and $\psi_{j\sigma}(x)$ being the creation and annihilation operators for an electron of spin $\sigma$ in region $j = s$ (\ssc) and $p$ (\psc). The BdG Hamiltonian for each region is given by 
\begin{subequations}
\begin{align}
\mathcal{H}_s &= \left(-\frac{\hbar^2\partial_x^2}{2 m} - \mu_s\right) \tau_z + \Delta_s \tau_x, \\
\mathcal{H}_p &= \left(-\frac{\hbar^2\partial_x^2}{2 m} - \mu_p\right) \tau_z + V_Z\sigma_z - i\Delta_p \partial_x \tau_x \sigma_x,
\end{align}
\end{subequations}
where $m$ is the electron effective mass, $\mu_{s}$ ($\mu_p$) is the chemical potential of \ssc (\psc), $V_Z$ is a Zeeman splitting, $\Delta_{s}$ ($\Delta_{p}$) is the \ssc (\psc) pairing potential, and $\tau_{x,y,z}$ ($\sigma_{x,y,z}$) are Pauli matrices acting in the particle-hole (spin) subspace. The effective chemical potential of each spin channel in the \psc ($\mu_p \pm V_Z$) is the key parameter determining whether that channel is topological or not. If the chemical potential of the channel is positive then the channel is topologically nontrivial, otherwise it is topologically trivial~\cite{Kitaev,Read}. In particular, we are interested in the regime where $|V_Z| > \mu_p$, in which \emph{one} of the channels is topologically trivial and the other is topologically nontrivial, leaving a single unpaired MZM at the end.

Figures~\ref{fig1}(b) and 1(c) show the zero-temperature differential conductances $G$ plotted against the bias voltage $V$ for the junction in Fig.~\ref{fig1}(a) in the limit of (b) large and (c) small Zeeman field. In the limit of large Zeeman field $[(|V_Z| -\mu_p) \sim \mu_p]$, the spinful \psc becomes effectively a spinless \psc~\cite{Kitaev,Read}. Owing to the difference in the spin dependence of Andreev reflection for the \ssc and spinless \psc, MAR is totally suppressed and only single Andreev reflections are allowed for this type of junction~\cite{Yeyati}. Correspondingly, the current at low voltages is zero until the bias voltage reaches $\Delta_s$ where the incoming quasiparticle acquires enough energy to undergo a single Andreev reflection from the MZM. This manifests as a step jump in the conductance to a quantized value $G_M = (4 - \pi)2e^2/h$. This quantization is robust against the junction transparency, as shown in Fig.~\ref{fig1}(b). Away from the threshold voltages $eV = \pm\Delta_s$, the conductance decreases with decreasing junction transparency and can become negative for sufficiently low transparencies. At high voltage, the conductance shows another jump to a value approaching the Landauer conductance value $G_N$ at $eV = \Delta_s + \Delta_{\mathrm{topo}}$, where $\Delta_{\mathrm{topo}}$ is the gap in the \psc. This marks the transition between subgap and above-gap conductance. Our result for the spinful \psc in the large Zeeman limit is similar to the conductance of the spinless \psc in Ref.~\cite{Yeyati} which is calculated using the Green's function formalism. In the opposite limit of small Zeeman field $[(|V_Z| -\mu_p) \ll \mu_p]$, shown in Fig.~\ref{fig1}(c), $G(e|V| = \Delta_s)$ is no longer quantized for intermediate and large transparencies, and the conductance develops ``subharmonic gap structure" (SGS)~\cite{KBT,Averin,Bagwell} at specific values of voltages corresponding to the gaps in the \ssc and \psc. Near zero voltage, the conductance is strongly suppressed because  $s$SC allows only spin-singlet Andreev reflections, while the MZM favors spin-triplet Andreev reflections~\cite{jjhe,xinliu}. However, away from the zero voltage MAR processes are not totally suppressed by spin-selectivity and do contribute to the SGS.  Only in the tunneling limit, where the current arises only from single Andreev reflections, do we obtain $G(\pm \Delta_s) = G_M$. Finally, although not shown, we note that the conductance is in general asymmetric in $V$ except for sufficiently small transparencies when MAR are totally suppressed. The particle-hole asymmetry of the conductance was recently observed in the scanning tunneling microscopy experiment of a one dimensional $p$-wave superconducting chain~\cite{Ben}.

Next, we move on to a more physically realistic model for the topological superconductor: the SOCSW. This model can be realized by proximitizing a spin-orbit-coupled semiconductor nanowire with an \ssc, in the presence of a spin-splitting magnetic field~\cite{roman, oreg, jay10,alicea}. The BdG Hamiltonian for the SOCSW is
\begin{align}
\mathcal{H}_{\rm SOCSW} = &\left(-\frac{\hbar^2\partial_x^2}{2m}-\mu_0\right)\tau_z -i \alpha \partial_x\tau_z\sigma_y + V_Z\sigma_x + \Delta_0\tau_x, \label{eq:HamSOCSW}
\end{align}
where  $\mu_0$ is the chemical potential, $\alpha$ is the spin-orbit coupling strength, $V_Z$ is the Zeeman field, and $\Delta_0$ is the proximity-induced $s$-wave pairing potential.

\begin{figure}[h!]
\capstart
\begin{center}
\includegraphics[width=\linewidth]{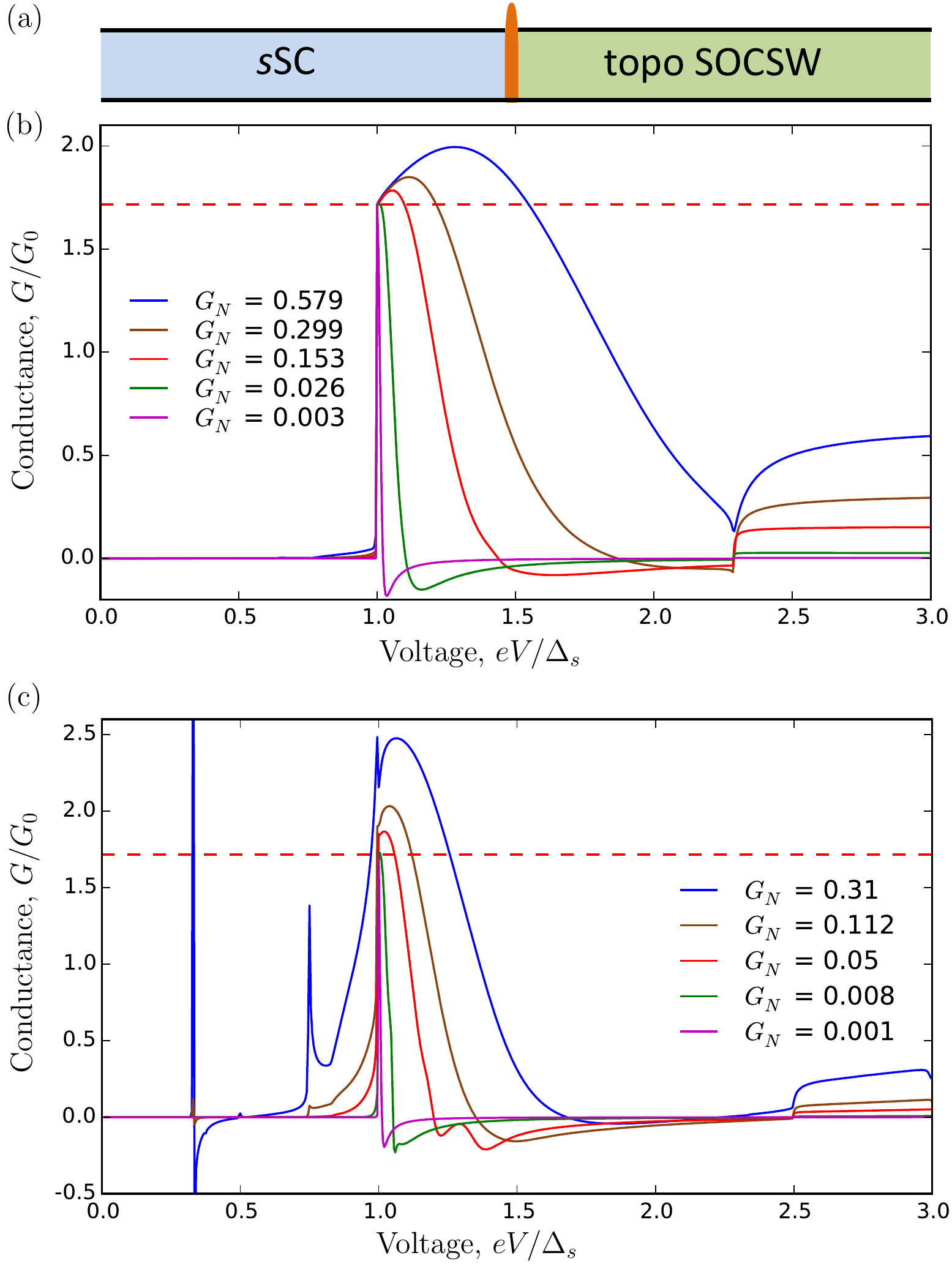}
\end{center}
\caption{(color online) (a) Schematic diagram for an \ssc-SOCSW junction with a delta barrier in the middle. (b),(c) Normalized differential conductance $G/G_0$ vs bias voltage $V$ for an \ssc-SOCSW junction with different Zeeman fields: (b) $V_Z$ = 15 K and (c) $V_Z$ = 1.65 K. We show the results for several transparencies $G_N$ where $G_N$ is the differential conductance at high voltages which is the conductance of the corresponding NN junction. All conductances are expressed in units of $G_0 = e^2/h$. The red dashed line [$G_M= (4-\pi)2e^2/h$] is the conductance value of tunneling into the MZM. In the tunneling limit, the conductance displays a step jump from $G= 0$ to $G = G_M$ at the threshold voltage $eV = \Delta_s$. In all plots, we used the following parameters: $m =$ 0.015$m_e$, $\mu_0 =$ 0 K, $\Delta_0$ = 1.5 K, $\mu_s =$ 200 K, $\alpha$ = 0.025 eV \AA, and $\Delta_s$ = 0.1 K. The topological gaps of the SOCSW are (b) $\Delta_{\mathrm{topo}} = 0.13$ K and (c) $\Delta_{\mathrm{topo}} =  0.15$ K.} \label{fig2}
\end{figure}

We calculate the conductance of an \ssc-SOCSW junction, shown schematically in Fig.~\ref{fig2}(a), where the SOCSW is in the topological regime (i.e., $V_Z > \sqrt{\mu_0^2 + \Delta_0^2}$~\cite{jay09,roman,oreg,jay10}). Similar to the \psc case, in the absence of MAR, there is a step jump in the conductance from $G = 0$ to $G = G_M$ as shown in Fig.~\ref{fig2}(b). This case, however, is not generic except in the tunneling limit. In general, for intermediate or large transparencies, MAR is present. As a result, the conductance $G(\pm\Delta_s)$ is not quantized, and there is SGS in the conductance profile [see Fig.~\ref{fig2}(c)]. As for the $p$SC case, the conductance near zero voltage is strongly suppressed due to the difference between the Andreev-reflection spin selectivity of the MZM and $s$SC. We note also that the same conclusion holds when the \ssc is replaced by an SOCSW in the nontopological regime (i.e., $V_Z < \sqrt{\mu_0^2 + \Delta_0^2}$).

\begin{figure}[h!]
\capstart
\begin{center}
\includegraphics[width=\linewidth]{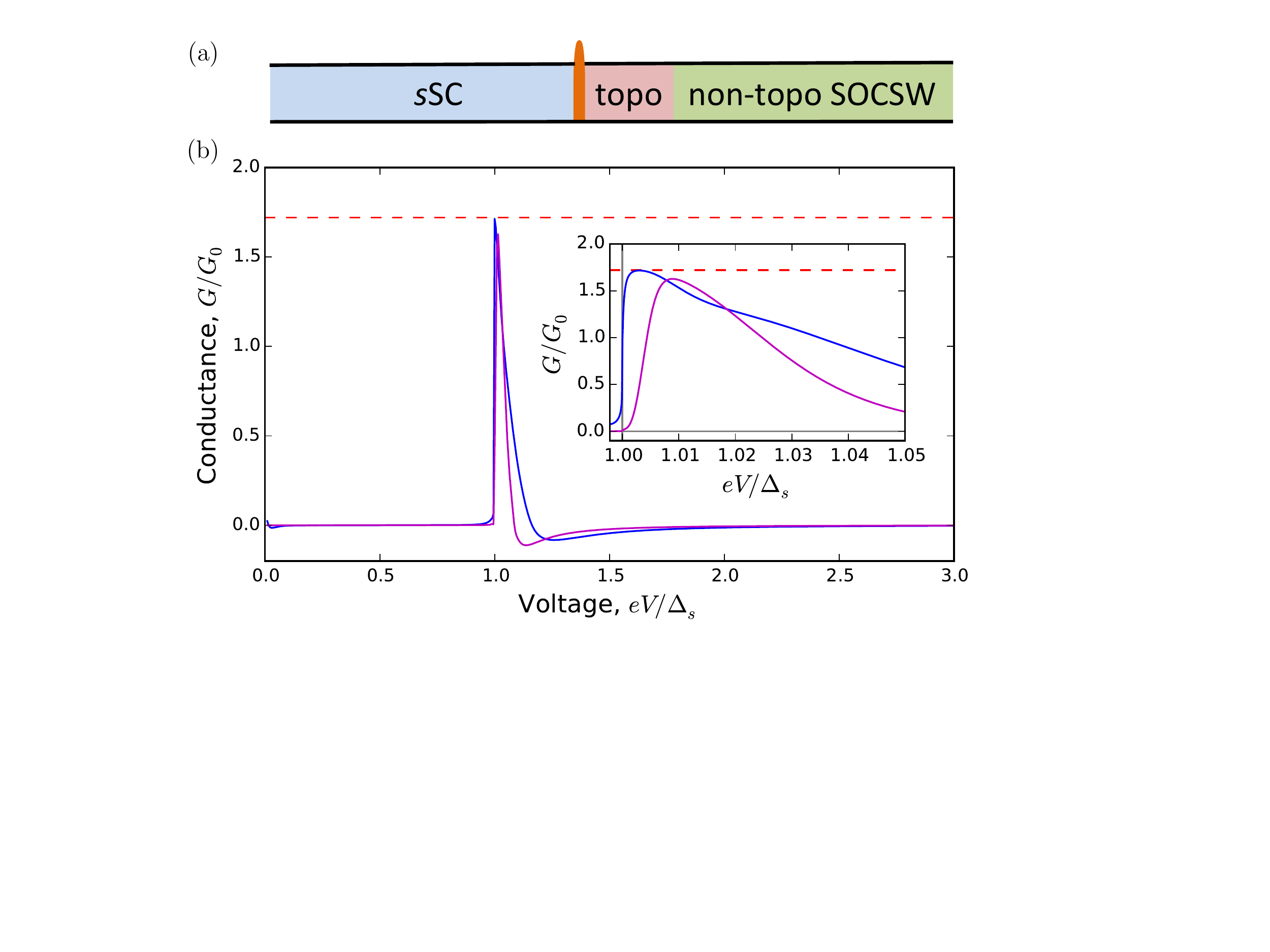}
\end{center}
\caption{(color online) (a) Schematic diagram for an \ssc-SOCSW junction with a pair of zero-energy ABS (one at each end of the topological region). The topological and nontopological regions have chemical potentials $|\mu_{\mathrm{topo}}| < \sqrt{V_Z^2 - \Delta_0^2}$ and $|\mu_{\mathrm{nontopo}}| > \sqrt{V_Z^2 - \Delta_0^2}$, respectively. The zero-energy ABS occurs at a specific value of $\mu_{\mathrm{topo}}$. (b) Normalized tunneling differential conductance $G/G_0$ vs bias voltage $V$ for the junction in (a) for different parameters: (i) blue curve: $\mu_{\mathrm{topo}} =$ 1.697 K, $\mu_{\mathrm{nontopo}} =$ 211.18 K, $V_Z =$ 15 K, $\Delta_0 =$ 10 K, and $\alpha =$ 0.025 eV$\cdot$\AA; (ii) magenta curve: $\mu_{\mathrm{topo}} =$ 0.9524 K, $\mu_{\mathrm{nontopo}} = $ 53.873 K, $V_Z =$ 4 K, $\Delta_0 =$ 1 K and $\alpha =$ 0.25 eV$\cdot$\AA. In all plots, we use $m = 0.015m_e$, $\mu_s = 200$ K, $\Delta_s =$ 0.1 K, and length of the topological region, $L_{\mathrm{topo}} = 0.6$ $\mu$m. All conductances are expressed in units of $G_0 = e^2/h$. For certain values of parameter (blue curve), the zero-energy ABS tunneling conductance may look like the Majorana conductance, i.e., it has a steep rise to the quantized value $G_M$ (red dashed line) near $eV = \Delta_s$. Inset: The zoom-in version of the conductance plot near the threshold voltage $eV = \Delta_s$.}\label{fig3}
\end{figure}

Finally, it is of interest to compare the conductance due to a single Andreev reflection from an MZM with that from a zero-energy but otherwise conventional ABS (e.g., a zero-energy Shiba state, or ``accidentally" unpaired Majorana doublet). In the following, we focus on the zero-energy ABS coming from an unpaired Majorana doublet which can arise in a superconductor having a finite topological region and semi-infinite nontopological region as shown in Fig.~\ref{fig3}(a).  One scenario where a nontopological zero-energy mode is naturally created 
near the end of an SOCSW is when the chemical potential $(\mu_0)$ of the wire in a high-density nontopological regime (i.e., $\mu_0>\sqrt{V_Z^2-\Delta_0^2}$) is depleted towards the end. If the potential is sufficiently smooth, then the chemical potential in going from the nontopological positive value, $\mu_0>\sqrt{V_Z^2-\Delta_0^2}$,  to negative values $\mu_0<-\sqrt{V_Z^2-\Delta_0^2}$ necessarily 
crosses the range of chemical potentials $|\mu_0|<\sqrt{V_Z^2-\Delta_0^2}$ where the SOCSW would be topological. The resulting domain walls between the topological segments and the nontopological SOCSW in the bulk can lead to a pair of zero-energy modes; one of which is closer 
to one of the leads~\cite{kell}. For simplicity, we consider a step jump of the chemical potential in going from the topological ($|\mu_{\mathrm{topo}}| < \sqrt{V_Z^2 - \Delta_0^2}$) to the nontopological regions ($|\mu_{\mathrm{nontopo}}| > \sqrt{V_Z^2 - \Delta_0^2}$)  with other parameters being the same. In this model, the ABS energy oscillates with chemical potential and Zeeman field, such that the zero-energy ABS occurs at specific values of parameters~\cite{Jay12}. We calculate the SNS conductance profile associated with this zero-energy ABS. As shown in Fig.~\ref{fig3}(b), in the tunneling limit the ABS conductance has a smooth onset rise at $eV = \Delta_s$ which develops into a non-quantized peak and then decreases. However, for certain parameter values [see blue curve in Fig.~\ref{fig3}(b)], the SNS conductance for the ABS may look similar to that of an MZM, i.e., it has a steep rise to the quantized value $G = G_M$ at $eV = \Delta_s$. Within experimental resolution, this zero-energy ABS peak may not be distinguishable from that of an MZM, making it difficult to use the SNS conductance quantization as an unambiguous signature for the MZM. The corresponding NS tunneling conductance peak for this particular ABS is $\sim1.6e^2/h$, which is significantly less than the quantized value of $2e^2/h$ for the MZM.

In conclusion, we have shown that in the presence of MAR, which is generic when the junction transparency is not small, the SNS conductance for a nontopological-topological superconductor junction is not quantized at the threshold voltages $e|V| = \Delta_s$, i.e., when the superconducting gap singularity lines up with the MZM. We have also shown that for some parameter values, the conductance of zero-energy ABS may look very similar to that of Majorana, such that the two cases may not be distinguishable within experimental resolution. This implies that, despite other benefits of using SNS junctions to probe MZMs, conductance quantization may not be a robust and definitive experimental signature.

Our theory actually has three important aspects which should be emphasized: (1) We provide a general theory within the scattering matrix formalism for transport in SNS junctions where one or both of the superconductors could be topological without making any approximation for weak tunneling and/or the number of Andreev bound states in the system. Our formalism can be used for any kind of SNS junctions requiring only NS scattering matrices which can be easily computed using Kwant~\cite{kwant}. Our formalism thus complements the Green's function method which is used to treat the topological superconductor junctions~\cite{Peng15,Yeyati,aguado,Denis}; (2) our theory shows that the  conductance in such topological junctions could be quite complex depending on the system parameters and any signature for Majorana zero modes are inherently subtle requiring a careful interpretation of the conductance using our theory; and (3) a necessary corollary of the last item is that the conductance quantization found earlier in the weak-tunneling limit of topological superconductor junctions~\cite{Peng15} is unlikely to be present in the generic experimental situation where the constraints of weak tunneling and/or number of Andreev bound states cannot be \textit{a priori} guaranteed.  Our theory should serve as the benchmark for future SNS conductance experiments and simulations where at least one of the superconductors is a topological superconductor.  It must also be emphasized that our work is completely general and applies to simple model systems of ideal spinless $p$-wave superconductors or spin-orbit-coupled nanowires in the presence of superconducting proximity effect and finite magnetic fields.

This work is supported by Microsoft Station Q, LPS-MPO-CMTC, and JQI-NSF-PFC. We acknowledge the University of Maryland supercomputing resources~\cite{hpcc} made available in conducting the research reported in this Rapid Communication.

\onecolumngrid
\vspace{1cm}
\begin{center}
{\bf\large Supplemental material for ``Conductance spectroscopy of nontopological-topological superconductor junctions"}
\end{center}
\vspace{0.5cm}

\setcounter{secnumdepth}{3}
\setcounter{equation}{0}
\setcounter{figure}{0}
\renewcommand{\theequation}{S-\arabic{equation}}
\renewcommand{\thefigure}{S\arabic{figure}}
\renewcommand\figurename{Supplementary Figure}
\renewcommand\tablename{Supplementary Table}
\newcommand\Scite[1]{[S\citealp{#1}]}
\newcommand\Scit[1]{S\citealp{#1}}

\makeatletter \renewcommand\@biblabel[1]{[S#1]} \makeatother

\section{Scattering matrix Formalism}

\begin{figure}[h!]
\capstart
\begin{center}\label{figsuppl1}
\includegraphics[scale = 0.6]{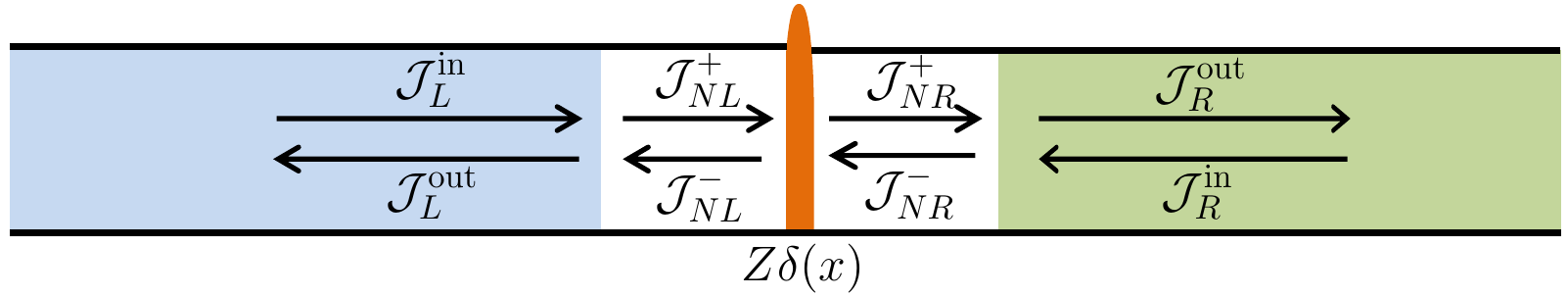}
\end{center}
\caption{Schematic diagram of a superconductor-normal metal-superconductor junction.} 
\end{figure}
We consider a  superconductor-normal metal-superconductor junction with a delta function tunnel barrier of strength $Z$ as depicted in Fig.~\ref{figsuppl1}. We assume that the normal region is infinitesimally short with a large chemical potential so that propagating modes in this region have constant group velocity. A quasiparticle can be injected either from the left or right superconductor. The incident quasiparticle from the left superconductor is transmitted as an electron or hole in the left normal region. In a voltage-biased junction, an electron (hole) gains (loses) an energy of $eV$ as it goes from the left to the right NS interface while at the NS interfaces, an electron or hole undergoes either normal or Andreev reflection. As a result, an electron coming from the left normal region at energy $E$ is Andreev reflected back into the same region as a hole with an energy $E+2eV$. These reflections happen repeatedly inside the normal region until the electron gains sufficient energy to transmit into one of the unoccupied bands in the superconducting leads. These repeated reflections are referred to as multiple Andreev reflections (MAR)~\cite{KBT,Averin,Bagwell}. The scattering processes for this junction can be separated into three parts: at the left NS interface ($S_L$), at the tunnel barrier ($S_N$) and at the right NS interface ($S_R$). More formally, it can be written as
\begin{subequations}\label{eq:matchingeq}
\begin{align}
\left( \bmat
\mathcal{J}_{L,\nu}^{\mathrm{out}} (E_n) \\
\mathcal{J}_{NL,\nu}^{+} (E_n)
\emat\right) = S_L(E_n) \left(\bmat \mathcal{J}_{L,\nu}^{\mathrm{in}} (E_n)\delta_{n0}\delta_{\nu,\rightarrow} \\\mathcal{J}_{NL,\nu}^{-} (E_n)\emat  \right) ,\\
\left( \bmat
\mathcal{J}_{NL,\nu}^{-} (E_n) \\
\mathcal{J}_{NR,\nu}^{+} (E_n) 
\emat\right) = \sum_{n'} S_N(E_n,E_{n'}) \left( \bmat \mathcal{J}_{NL,\nu}^{+} (E_{n'}) \\ \mathcal{J}_{NR,\nu}^{-} (E_{n'}) \emat\right),\\
\left( \bmat
\mathcal{J}_{R,\nu}^{\mathrm{out}} (E_n)\\
\mathcal{J}_{NR,\nu}^{-} (E_n) 
\emat\right) = S_R(E_n) \left(\bmat \mathcal{J}_{R,\nu}^{ \mathrm{in}} (E_n)\delta_{n0}\delta_{\nu,\leftarrow} \\ \mathcal{J}_{NR,\nu}^{+} (E_n) \emat  \right),
\end{align}
\end{subequations} 
where $E_n = E+ neV$ with $n$ being an integer, $\mathcal{J}_{\ell,\nu}^\rho = (j_{\ell,\nu}^{e,\uparrow,\rho},j_{\ell,\nu}^{e,\downarrow,\rho},j_{\ell,\nu}^{h,\uparrow,\rho},j_{\ell,\nu}^{h,\downarrow,\rho})^{\mathrm{T}}$ is the current amplitude vector for region $\ell = L$ (left superconductor), $NL$ (normal region to the left of the tunnel barrier), $NR$ (normal region to the right of the tunnel barrier) and $R$ (right superconductor) with $\rho = +/-$ and $\rho = \mathrm{in}/\mathrm{out}$ being the indices for the right/left-moving and incoming/outgoing modes, respectively. The index $\nu = \rightleftarrows$ denotes whether the injected current is from the left or right superconductor. The scattering matrix $S_N(E_n,E_n')$ takes into account the fact that the energy of an electron (hole) increases (decreases) by $eV$ every time it passes from left to right and also the normal reflection by the delta function tunnel barrier. It can be decomposed into the electron ($S_{N}^e$) and hole ($S_{N}^h$) part:
\beq
S_N(E_n,E_{n'}) = S_{N}^e(E_n,E_{n'})\otimes \sigma_0 \otimes \tau_+ + S_{N}^h(E_n,E_{n'}) \otimes \sigma_0\otimes\tau_-,
\eeq
where $\sigma_0$ is the identity in spin subspace, $\tau_\pm = \tau_x \pm i\tau_y$ are the Pauli matrices in particle-hole subspace,
\begin{align}
S_{N}^e(E_n,E_{n'}) &= \left(\bmat r \delta_{n,n'} & t \delta_{n,n'+1} \\ t \delta_{n,n'-1} & r \delta_{n,n'} \emat \right), \nl
S_{N}^h(E_n,E_{n'}) &= \left(\bmat r^* \delta_{n,n'} & t^* \delta_{n,n'-1} \\ t^* \delta_{n,n'+1} & r^* \delta_{n,n'} \emat \right),
\end{align}
with the reflection coefficient $r = -iZ/(1+iZ)$ and transmission coefficient $t = 1/(1+iZ)$ being dependent on the strength of the delta barrier $Z$.

By solving Eq.~\eqref{eq:matchingeq} for the current amplitudes $\mathcal{J}^{\pm}_{NL}$ with incident quasiparticle from the left and right superconductor ($\nu = \rightleftarrows$), we can obtain the zero-temperature dc-current $I(V) = I_{\rightarrow}(V) + I_{\leftarrow}(V)$ from
\begin{align}\label{eq:curr}
I_{\nu}(V) = \frac{2 e}{h}\int_{-\infty}^{0} dE \mathrm{Tr}\left(\sum_{n} \rho_z \tau_z J_{NL,\nu}(E_n) J_{NL,\nu}^\dagger(E_n)\right),
\end{align}
where $J_{NL,\nu} = (j_{NL,\nu}^{e,\uparrow,+},j_{NL,\nu}^{e,\downarrow,+}, j_{NL,\nu}^{h,\uparrow,+},j_{NL,\nu}^{h,\downarrow,+}, j_{NL,\nu}^{e,\uparrow,-},j^{e,\downarrow,-}_{NL,\nu}, j_{NL,\nu}^{h,\uparrow,-},j_{NL,\nu}^{h,\downarrow,-})^{\mathrm{T}}$ is the current amplitude vector in the normal region to the left of the barrier. The differential conductance ($G = dI/dV$) is calculated by simply differentiating the current $I$ with respect to the voltage $V$.

\section{Remarks on Numerical Simulation}
We obtained the scattering matrices at the left ($S_L$) and right NS interfaces ($S_R$) [Eq.~\eqref{eq:matchingeq}] from Kwant~\cite{kwant} by setting up the tight-binding models for the corresponding NS junctions. Since Kwant chooses arbitrary phases for the propagating modes at each energy, we fixed the phases of the propagating modes by setting the largest element of propagating modes for every energy to be real. 

We note that Eqs.~\eqref{eq:matchingeq}(a) and (c) are invariant under the transformation
\begin{align}
t_{L,R}^{\mathrm{in}}(E) \rightarrow t_{L,R}^{\mathrm{in}} (E) U_{L,R}^\dagger (E), \nonumber\\
\mathcal{J}_{L,R}^{\mathrm{in}}(E) \rightarrow U_{L,R}(E)\mathcal{J}_{L,R}^{\mathrm{in}}(E),
\end{align}
where $t_{L,R}^{\mathrm{in}}(E)$ is the transmission matrices at the left and right NS interfaces, $\mathcal{J}_{L,R}^{\mathrm{in}}(E)$ are the input current amplitudes from the left and right NS interfaces, and $U_{L,R}(E)$ are unitary matrices. By polar decomposition, there exists a unitary matrix $U_{L,R}(E)$ such that $t_{L,R}^{\mathrm{in}}(E) = \widetilde{t}_{L,R}^{\mathrm{in}}(E) U_{L,R}^\dagger (E)$, where
\beq\label{eq:transmatrix}
\widetilde{t}_{L,R}^{\mathrm{in}}(E) = \sqrt{t_{L,R}^{\mathrm{in}}(E)[t_{L,R}^{\mathrm{in}}(E)]^\dagger} = \sqrt{\mathds{1}-r_{L,R}(E) r_{L,R}^\dagger(E)}, 
\eeq
where $r_{L,R}$ are the reflection matrices at the left and right NS interfaces. To speed up the computation, we obtained only the reflection matrices $r_{L,R}$ from Kwant and used Eq.~\eqref{eq:transmatrix} to calculate the transmission matrix.

Numerically, we introduced an energy cutoff $E_c$ in the summation over energy in Eq.~\eqref{eq:curr} where $E_c$ is chosen so that the calculation converges for each value of $V$. To ensure that the scattering matrix remains unitary after introducing the energy cutoff, we impose the following constraint on the scattering matrix:
\begin{align}
S_{N}^e(E,E+eV) = S_{N}^h(-E,-(E+eV)) = -\mathds{1},
\end{align}
for all $E > E_c$. 

\section{Proof for the non-negativity of the current}
The current amplitude in the normal region is given by
\begin{align}\label{eq:inpcurr}
\widetilde{j}^{\mathrm{tot}}_{\ell,\nu}(E) &= \widetilde{j}^{e}_{\ell,\nu}(E)+ \widetilde{j}^{h}_{\ell,\nu}(E),
\end{align}
where 
\begin{equation}
\widetilde{j}_{\ell,\nu}^\tau(E)  = \sum_{\sigma=\uparrow,\downarrow} j_{\ell,\nu}^{\tau,\sigma,+}(E) - j_{\ell,\nu}^{\tau,\sigma,-}(E),
\end{equation}
is the electron/hole ($\tau = e/h$) component of the current in the left ($\ell = NL$) or right ($\ell = NR$) normal region. Since the electron (hole) energy increases (decreases) by $eV$ every time it passes from the left to the right, we have
\begin{subequations}\label{eq:lrcurr}
\begin{align}
\jtle(E) &= \jtre(E+eV), \\
\jtlh(E) &= \jtrh(E-eV).
\end{align}
\end{subequations} 
From Eqs.~\eqref{eq:inpcurr} and \eqref{eq:lrcurr}, we obtain the following recurrence relation
\begin{align}
\jtle(E) = \iinpl(E) - \iinpr(E-eV) + \jtle(E-2eV),
\end{align}
which implies that
\beq\label{eq:recur}
\jtle(E) = \sum_{n=0}^{\infty}\iinpl(E-2neV) - \iinpr(E-(2n+1)eV).
\eeq
The total electrical current is given by
\begin{equation}\label{eq:totcurr}
I_{\nu} = \frac{2e}{h} \int dE \sum_{m} \left[\jtle(E+2meV) - \jtlh(E+2meV)\right].
\end{equation}
Multiplying the integrand of Eq.~\eqref{eq:totcurr} by $V$, we have the power dissipated by the SNS junction as
\begin{align}\label{eq:positivecurr}
I_\nu V &= \frac{2eV}{h}\int dE \sum_{m}  \left[2\jtle(E + 2meV) - \iinpl(E+ 2meV)\right],\nonumber\\
&= \frac{2eV}{h}\int dE \sum_{m} \left\{ 2\sum_{n\geq 0} \left[\iinpl (E + 2(m - n)eV)-\iinpr(E+(2(m-n)-1)eV)  \right] - \iinpl(E+2meV)\right\}, \nonumber\\
&= \frac{2eV}{h} \int dE \sum_{m} \left\{ 2\sum_{m'\leq m} \left[ \iinpl(E + 2m'eV) - \iinpr(E + (2m'-1)eV)\right]-\iinpl(E + 2meV)\right\}, \nonumber\\
&= \frac{2eV}{h} \int dE \left\{  2\sum_{m'} (m_{\mathrm{max}} -m'+1)\left[ \iinpl(E + 2m'eV) - \iinpr(E + (2m'-1)eV)\right]-\sum_{m}\iinpl(E+ 2meV) \right\},  \nonumber\\
&= \frac{2eV}{h} \int dE \sum_{m} \left[(2m-1)\iinpr(E+(2m-1)eV) -2m\iinpl(E+2meV) \right] \geq 0.
\end{align}
So, for $V \geq 0$, we have $I_{\nu} \geq 0$. In lines 1 and 2, we have made use of Eqs.~\eqref{eq:inpcurr} and \eqref{eq:recur}, respectively. In lines 4 and 5 of Eq.~\eqref{eq:positivecurr}, we have used the current conservation equation:
\beq
\sum_m \left[\iinpl(E+2meV) - \iinpr(E+(2m-1)eV)\right] = 0,
\eeq 
and the fact that $\pm(E+neV)\widetilde{j}^{\mathrm{tot}}_{NL/NR,\nu}(E+neV) \geq 0$ is the power dissipated by current $\widetilde{j}^{\mathrm{tot}}_{NL/NR,\nu}$.

\end{document}